\documentclass[a4paper]{jpconf}
\usepackage{graphicx}
\usepackage{natbib}

\newcommand{\Teff}{\hbox{$T\sb{\rm eff}$}}
\newcommand{\logg}{\hbox{$\log g$}}
\newcommand{\Msun}{\hbox{M$\sb{\odot}$}}

\def\aj{AJ}                   % Astronomical Journal
\def\apj{ApJ}                 % Astrophysical Journal
\def\apjl{ApJ}                % Astrophysical Journal, Letters
\def\apjs{ApJS}               % Astrophysical Journal, Supplement
\def\aap{A\&A}                % Astronomy and Astrophysics
\def\mnras{MNRAS}             % Monthly Notices of the RAS
\def\jrasc{JRASC}             % Journal of the RAS of Canada

\begin{document}
\title{SDSS White Dwarf mass distribution at low effective temperatures}

\author{D. Koester$^1$, S.O. Kepler$^2$, S.J. Kleinman$^3$ and A. Nitta$^3$}

\address{$^1$ Institut f\"ur Theoretische Physik und Astrophysik,
  Universit\"at Kiel, 24098 Kiel, Germany}
\address{$^2$ Instituto de Fisica, Universidade Federal do Rio Grande
  do Sul, 91501-970 Porto Alegre, RS Brazil}
\address{$^{3,4}$ Gemini Observatory, 670 N A'ohoku Pl., Hilo HI 96720 USA}

\ead{koester@astrophysik.uni-kiel.de}

\begin{abstract}
The DA white dwarfs in the Sloan Digital Sky Survey, as analyzed in the
papers for Data Releases 1 and 4, show an increase in surface gravity
towards lower effective temperatures below 11500~K. We study the
various possible explanations of this effect, from a real increase of
the masses to uncertainties or deficiencies of the atmospheric
models. No definite answer is found but the tentative conclusion is
that it is most likely the current description of convection in the
framework of the mixing-length approximation, which leads to this effect.
\end{abstract}

\section{Introduction}
Almost every contribution at this white dwarf meeting mentions the
Sloan Digital Sky Survey (SDSS) or is even based on its data. Because
of the huge impact the survey has on this field, it is very important
to critically analyze all aspects of the data collection and reduction
process, to understand possible uncertainties and biases. One problem
related to white dwarfs, which was obvious already in the first white
dwarf catalogue publication \citep{Kleinman.Harris.ea04}, was the
apparent systematic increase of the mean surface gravity of DAs below
effective temperatures of approximately 12000~K, determined from
spectral fits, was again demonstrated in the data of the Data Release
4 \citep{Eisenstein.Liebert.ea06} and confirmed by
\citet{Kepler.Kleinman.ea07} and \citet{DeGennaro.von-Hippel.ea08}.

While the huge number of objects and the homogeneity of the survey
make this feature very obvious, it is by no means a new discovery. A
similar trend was first recognized with data from the Palomar-Green
Survey \citep[][and several other papers of the Montreal
group]{Bergeron.Wesemael.ea90, Bergeron92}, and more recently by
\citet{Liebert.Bergeron.ea05}. A similar effect was also found in the
data from the ESO Supernova Ia Progenitor Survey (SPY) by
\citet{Voss06}. A first conclusion therefore is that this effect is
not a peculiarity of the SDSS data collection or reduction
processes. In this paper we study the remaining possibilities
\begin{itemize}
\item the effect could be real, i.e. the DAs show a real increase of
  the average mass towards lower temperatures
\item it could be caused by the fitting between observations and
  theoretical models used to determine the atmospheric parameters
\item uncertainties in input physics or a flaw in the models used
\end{itemize}
For our study we use the data of DR4 as provided from the SDSS
database, with slight changes in reduction software compared to
\citet{Eisenstein.Liebert.ea06}.  We also use very nearly the same
models as used in that work for our ``standard analysis'' (see
below). The major difference are the fitting algorithms: whereas the
SDSS uses the whole spectra, adjusted with a polynomial using
information from the photometry, we here use only the spectral lines,
with the continuum adjusted on both sides of the line center.  This
technique minimizes the influence of uncertainties in the flux
calibration.  The SDSS photometry gives an independent estimate of
parameters, which is used to distinguish between the two solutions,
which exist for most objects between 8000 and 18000~K.

In the main part of this study we vary some of the parameters in the
input physics, in the hope of finding some hints for the origin of the
effect. This concerns the treatment of convection, of the non-ideal
effects in the occupation numbers, and a possible admixture of
invisible helium. All these possible explanations have been proposed
and studied in various papers of the Montreal group. We partially
repeat this work here with a much larger and very homogeneous sample.

\section{Statement of the problem}
\begin{figure}[ht]
\begin{minipage}[t]{7.7cm}
\includegraphics[width=7.7cm]{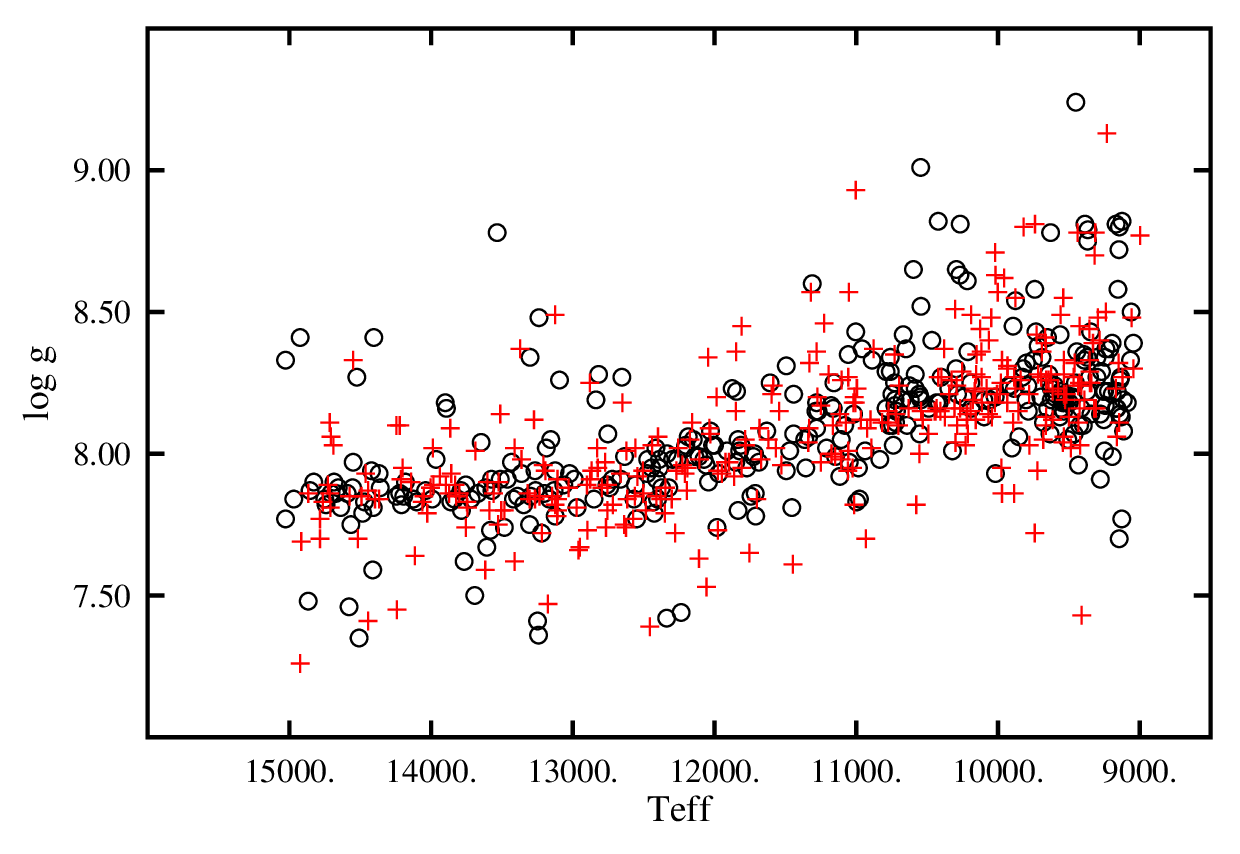}
\caption{\label{sdss-results} Distribution of \logg\ with \Teff\ for a
  subsample of cool DAs from DR4. Crosses (red): 347 DAs with
  magnitude $g < 18$; circles (black): 360 DAs with $g$ between 18 and
19.}
\end{minipage}\hspace{0.4cm}\begin{minipage}[t]{7.7cm}
\includegraphics[width=7.7cm]{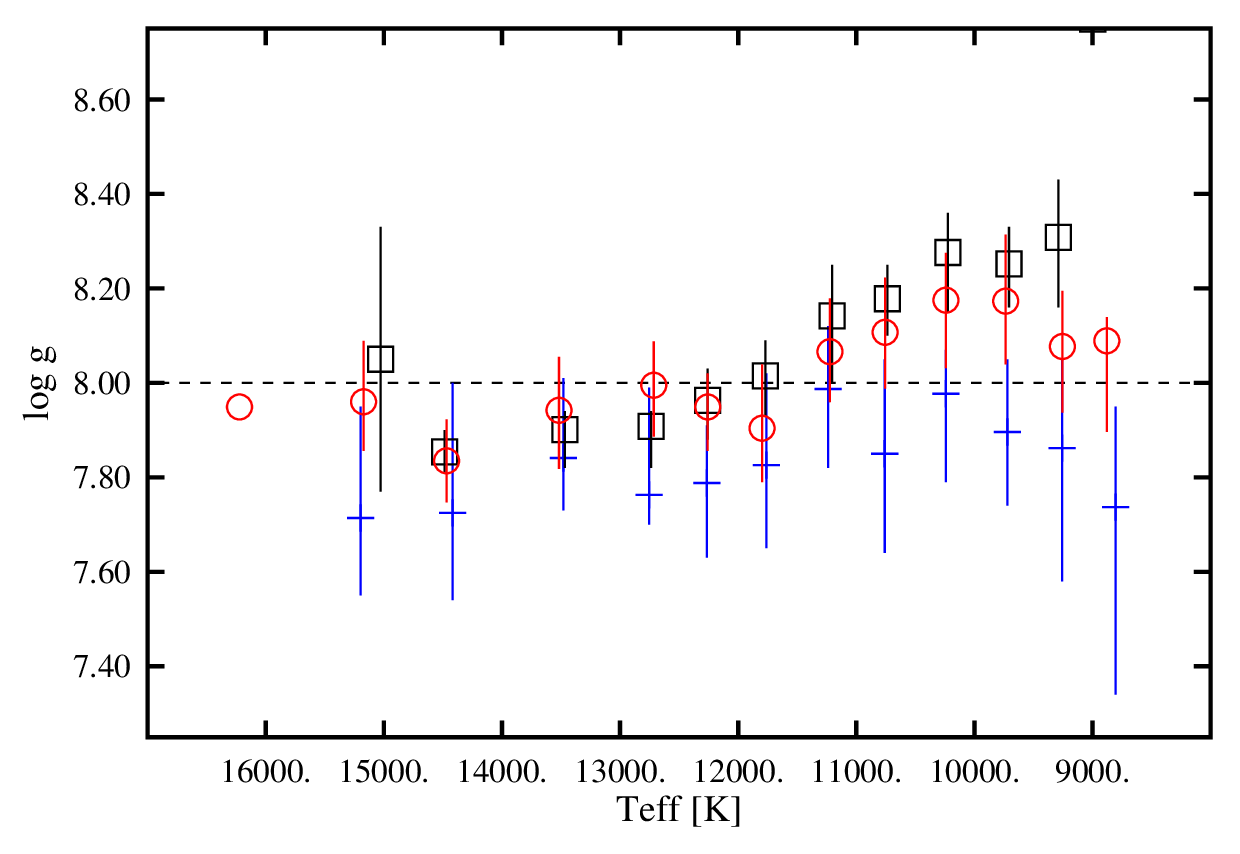}
\caption{\label{g-dist3} Average \logg\ values and distribution in bins
  of \Teff. Squares (black): original results from DR4; circles (red):
our new fits to the spectra; horizontal bars (blue): results from
photometry for the same objects. The vertical lines in all cases are
the intervals containing the ``inner'' 50\% of the distribution,
cutting 25\% on either side.}
\end{minipage} 
\end{figure}

Figure~\ref{sdss-results} shows a subsample of the brighter, cool
($<16000$~K) DAs in the DR4 results. The parameters were determined by
SJK; the only significant difference to
\citet{Eisenstein.Liebert.ea06} was a larger model grid.  While the
scatter of the distribution and trend towards large \logg\ is similar
for the two subsets, it is much smaller than in the total sample,
which includes many fainter stars (see DR4 for a comparison). A S/N of
at least 20, and preferably $>30$ is a minimum requirement for
reliable spectral analysis; we will therefore in the remaining tests
use the bright sample with $g < 19$. The trend in \logg\ becomes even
clearer, if we bin the results in \Teff\ intervals of 500 to 1000~K
(Fig.~\ref{g-dist3}). The spectroscopic fits show a clear increase
below \Teff\ = 11500~K, which is somewhat more pronounced in the DR4
results than in our own fits. Since everything else is very similar,
we attribute this difference to the different fitting methods. A
tentative explanation could be that the SDSS polynomial correction to
the flux calibration is problematic for the very extended wings of the
broad Balmer lines, especially in this temperature range.

The surface gravity derived from fitting the 5 SDSS magnitudes
simultaneously are slightly lower than the spectroscopic results. They
also show a small increase of \logg\ for lower \Teff\, but this is
much less pronounced than in the spectroscopic data. We would like to
make a few remarks regarding the SDSS photometry, which is not used as
much for the analysis of white dwarfs as it deserves. In the case of
the DAs the discrimination for \Teff\ and \logg\ is very good in the
range from 16000~K to 8000~K, and a determination using e.g. the
two-color diagram $u-g$ vs. $g-r$ can in principle give results as
accurate as from spectroscopy, without the problem of having two
possible solutions above and below the Balmer line maximum. However,
there is still some small uncertainty in the calibration of the
magnitude zero-points. As a test, we have increased the observed $g$
magnitude by 0.03, which results in a perfectly constant average
\logg\ sequence close to 8.0. While we do not imply that the $g$
uncertainty is as large as this, we believe that because of the
calibration uncertainties the photometry should at this time not have
very high weight in a study of {\em systematic} effects in the stellar
parameters. We will, however, in the following sections also show the
effect of model changes on the photometric solutions, which in an
ideal case would agree with the spectroscopic solution.

\section{Could the mass change with \Teff\ be real?}
The input physics for our ``standard'' models is the same as used for
the DR4 fits and will be described in more detail below. Our
spectroscopic fits with these models to the $\approx 700$ DAs brighter
than $g = 19$ and \Teff\ between 8500 and 16000~K is our reference set
for this study. There are 61 DAs with multiple spectra in this set;
from the differences in the determined parameters we estimate a
typical uncertainty of 2.4\% for \Teff, and 0.15 in \logg. Since the
instrumentation, reduction, models, and analysis methods were the
same, this is certainly still an underestimate of the real
errors. Nevertheless, the number of objects are large, and the
\logg\ increase so systematic that it certainly cannot be explained
by uncertainties of the parameter determinations.

With our reference fits the mean mass of the DAs is 0.573~\Msun\ for
the objects above 11500~K and 0.658~\Msun\ below. Could this increase
of the white dwarf masses be a real effect? Possible explanations
might be different cooling times for high versus low mass white
dwarfs, or higher mass progenitors (and consequently high mass wds)
for older stars, which originated from older progenitors with lower
metal abundances. The latter argument is, however, contradicted by the
findings of \citet{Hansen.Anderson.ea07}, that old white dwarfs in globular
clusters have masses close to 0.5~\Msun. All these arguments have been
discussed in \citet{Kepler.Kleinman.ea07} and found not to be
convincing. It is also very difficult to explain the rather sudden
increase around 11500~K with these explanations, which all should lead
to a more gradual transition if any.

\section{Non-ideal effects in equation of state and occupation
  numbers} The standard treatment of these effects in white dwarf
atmosphere models is the Hummer-Mihalas-D\"appen occupation
probability mechanism \citep{Hummer.Mihalas88, Mihalas.Dappen.ea88,
  Dappen.Mihalas.ea88}. For any effect above 10000~K only the
interactions with charged particles are of
interest.  \citet{Bergeron93} has demonstrated that a more consistent
fit between low and higher Balmer lines can be achieved by increasing
the parameter $\beta_{crit}$ in the HMD formalism from 1 to 2. The
effect is to decrease the absorption in the wings of the Balmer lines,
compared to the standard treatment. 
\begin{figure}[ht]
\begin{minipage}[t]{7.7cm}
\includegraphics[width=7.7cm]{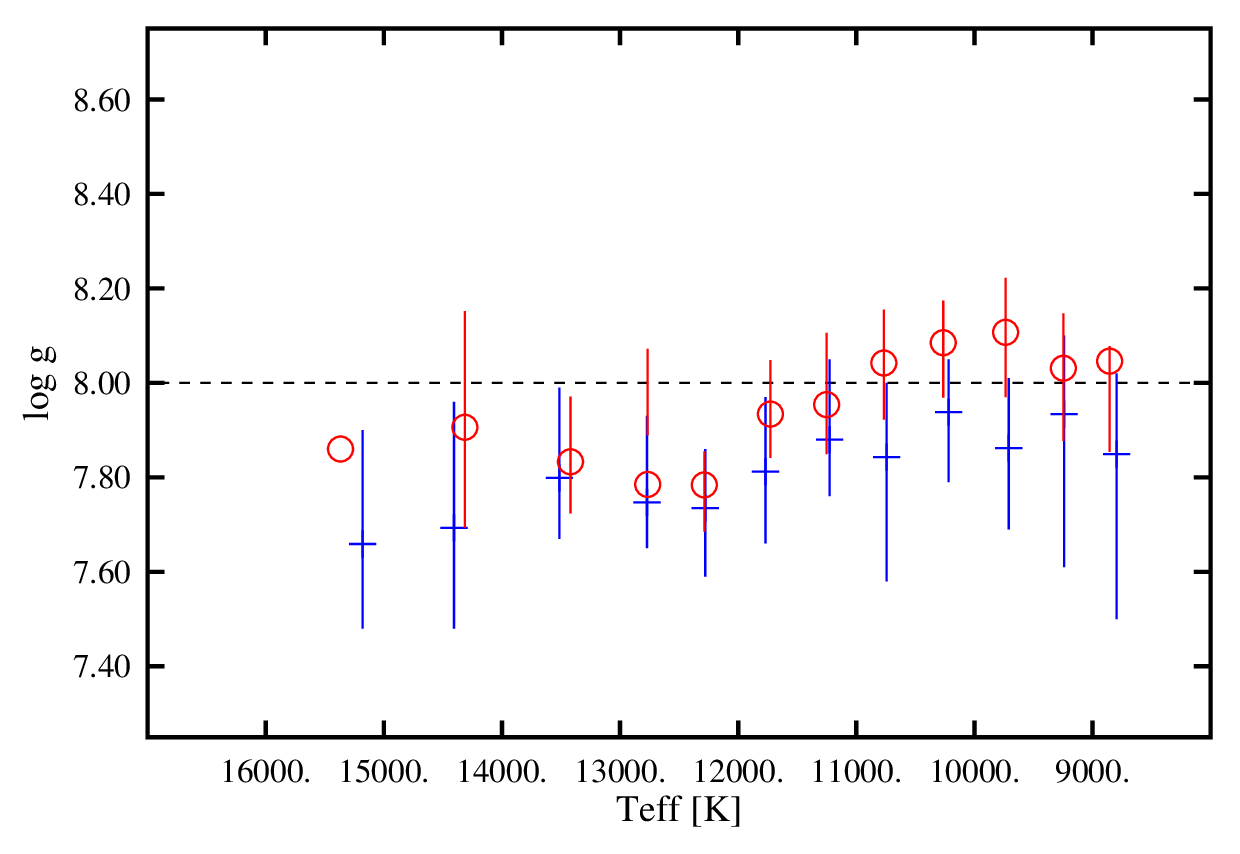}
\caption{\label{beta1} Average of \logg\ as a function of \Teff\  
  using models with $\beta_{crit}$ decreased to 1. Circles (red):
  spectroscopic solution; horizontal bars (blue): photometric solution}
\end{minipage}\hspace{0.4cm}\begin{minipage}[t]{7.7cm}
\includegraphics[width=7.7cm]{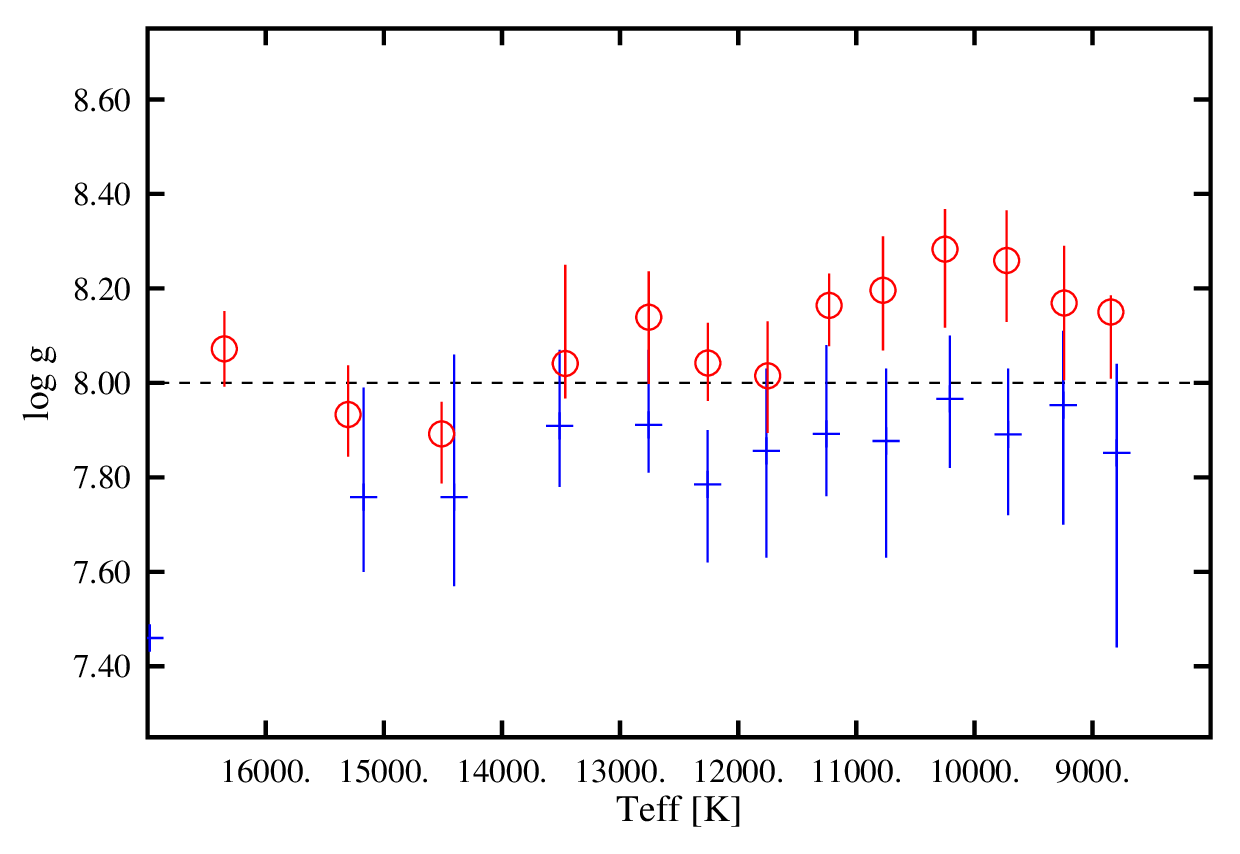}
\caption{\label{beta5} Similar to Fig.~\ref{beta1}, but $\beta_{crit}$
  increased to 5.}
\end{minipage} 
\end{figure}
We have experimented with this parameter and show two extreme cases in
Figs.~\ref{beta1} and \ref{beta5}. For the smaller value (stronger
line wings) the spectroscopic solution results in smaller \logg, as
expected, and the values agree better with the photometric
solution. The opposite is true for larger $\beta_{crit}$. Neither of
the changes offers any real improvements, since the jump at 11500~K
remains. This could also be expected, because the effect of changing
the strength of the charged particle interaction should be continuous
around this critical region. We therefore exclude this change as a
possible solution.

\section{Invisible helium in the atmosphere} 
This possible explanation was also first proposed by the Montreal
group \citep{Bergeron.Wesemael.ea90}. Why could this solve the problem
of increasing \logg? The surface gravity, which along with \Teff\ are
the stellar parameters determined in an analysis, is only a proxy. The
emission from a hot plasma is -- in addition to the elemental
abundances -- determined by the thermodynamic variables, e.g.
pressure and temperature. These variables are of course not constant
throughout the visible atmosphere, but they can be described by
characteristic values. For the temperature this is the effective
temperature. The pressure is determined by the hydrostatic equation
$   \frac{dP}{d\tau} = \frac{g}{\kappa} $
Assuming a constant absorption coefficient (per mass) $\kappa$ this
can be integrated to $\tau = 1$ to give $ P \sim g/\kappa$,
showing that the pressure is inversely proportional to the absorption
coefficient per mass. The latter is much smaller for helium than for
hydrogen around 10000 - 12000 ~K, and therefore a significant helium
pollution leads to higher pressure, mimicking a higher surface
gravity.

A possible source of helium could be helium dredge-up by the
convection zone (CVZ), which (with the standard description of
convection in white dwarfs) increases significantly in depth for
effective temperatures below 11500~K.
\begin{figure}[ht]
\includegraphics[width=7.7cm]{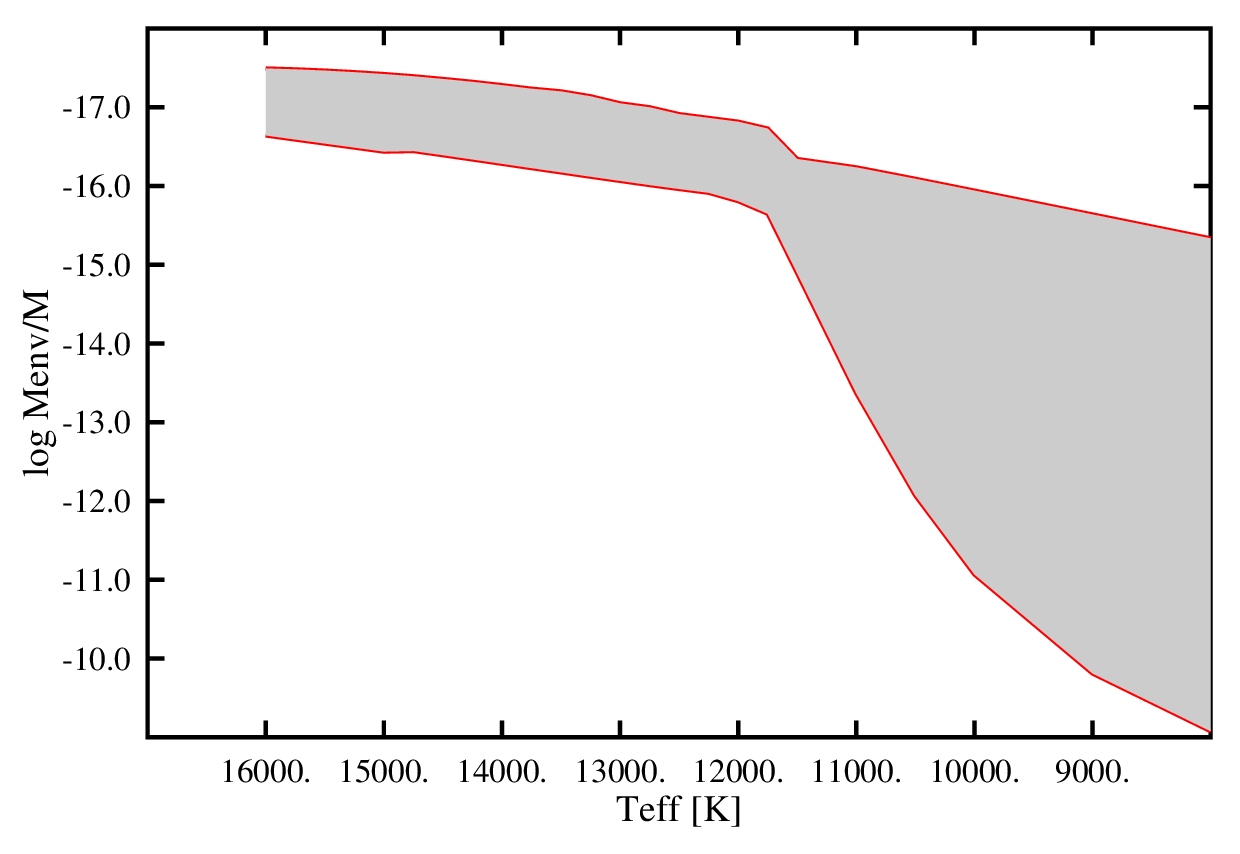}\hspace{0.4cm}%
\begin{minipage}[b]{7.7cm}\caption{\label{conv} Size of the convection
  zone in DA white dwarfs as a function of \Teff. The two continuous
  (red) lines are the upper and lower boundary, the gray area is the
  convection zone. The
  vertical axis is a mass scale, giving the logarithm of the fractional
  mass counted from the surface of the star. Note the sharp increase
  in depth around 11500~K. The mixing-length parameters for this
  calculation are ML2/0.6 (see text)}
\end{minipage}
\end{figure}
If the total amount of hydrogen in the outer layer is less than about
$10^{-10}$ of the total mass, such a mixing could happen near the
interesting temperature range.  

\begin{figure}[ht]
\begin{minipage}[t]{7.7cm}
\includegraphics[width=7.7cm]{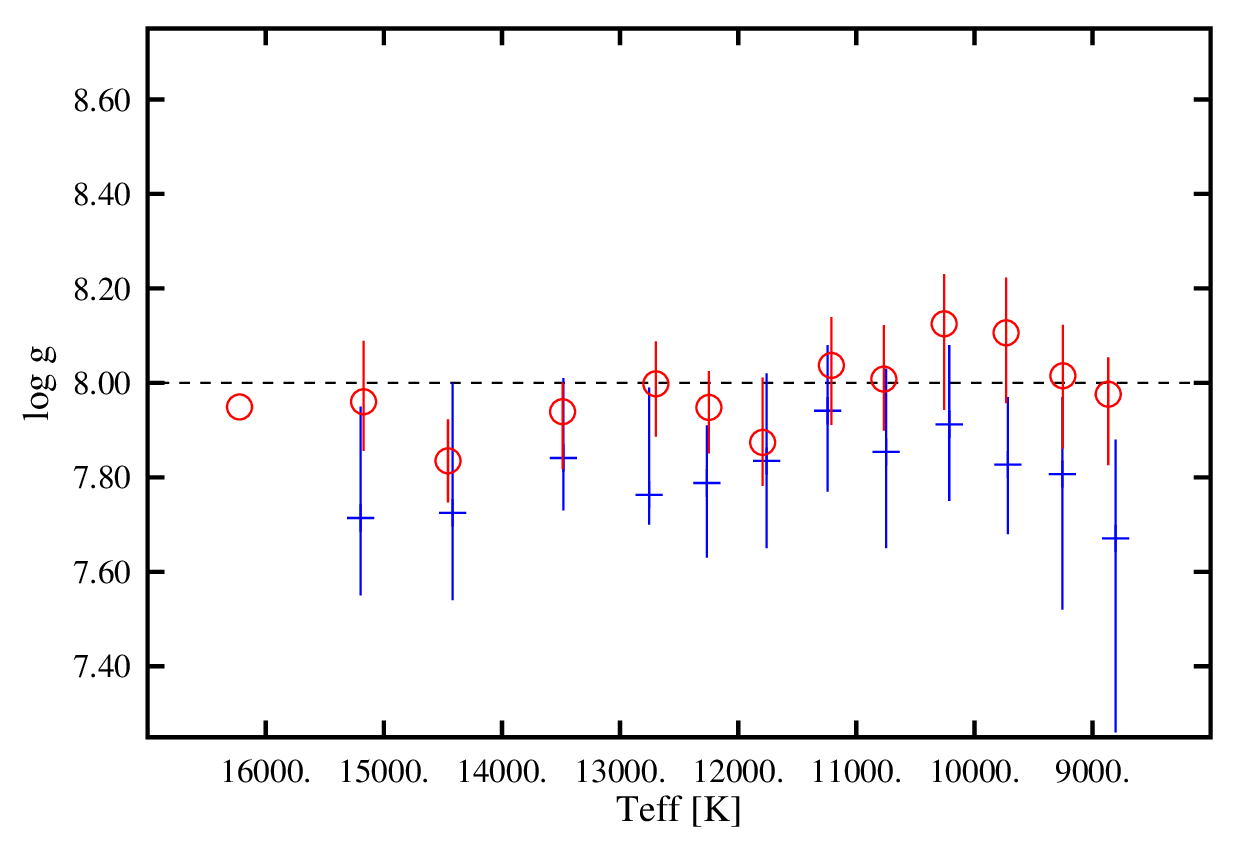}
\caption{\label{he01} Effect of adding helium to the atmosphere with
  a number ratio of He:H = 0.1. Symbols and colors are the same as in
  the previous figures.} 
\end{minipage}\hspace{0.4cm}\begin{minipage}[t]{7.7cm}
\includegraphics[width=7.7cm]{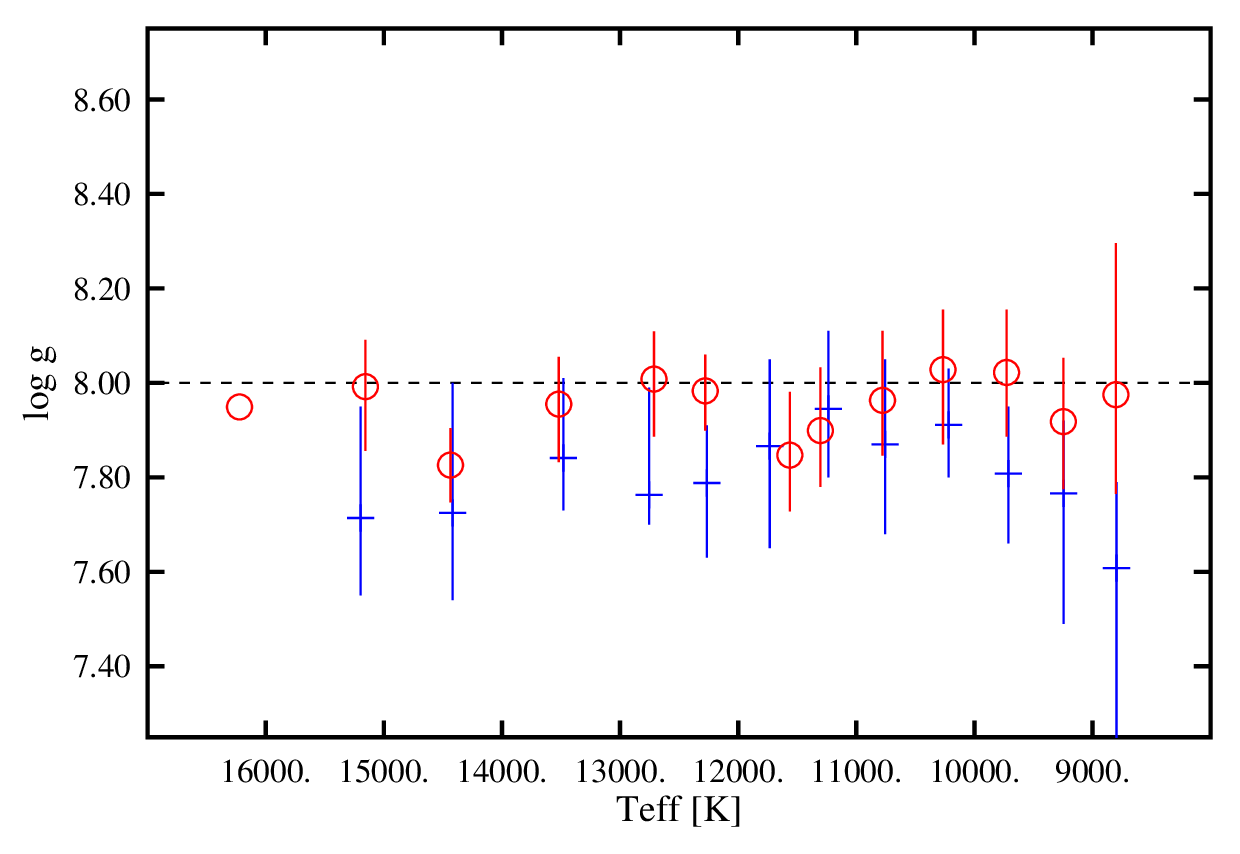}
\caption{\label{he10} Adding helium with an abundance of He:H = 1 by
  numbers.}
\end{minipage} 
\end{figure}
 
For simplicity we show in Figs.~\ref{he01} and \ref{he10} the effect
of adding a helium contamination with He:H = 0.1 and 1 by numbers in
all models with $\Teff \le 11500$~K. Obviously this goes in the
direction of a constant mass for all temperatures. In reality we would
expect a helium pollution, which increases with decreasing \Teff, with
an even more convincing result (removing the dip between 11000 and
12000~K).

Could such a helium contamination be detected directly? We remind the
reader of two recently discovered objects, which provide some guidance
- HS 0146+1847 and GD362. Both stars were at first classified as
massive DAZ, with very strong Balmer lines and CaII
\citep{Koester.Napiwotzki.ea05, Gianninas.Dufour.ea04,
  Zuckerman.Koester.ea07}. Yet, at closer inspection, they turned out
to be in fact helium-rich, with normal masses. In both cases high
resolution and high S/N spectra were necessary to discover the weak
helium lines, although the helium content is far larger than necessary
to explain the average \logg\ increase. In order to find more similar
objects we have systematically searched $\approx 1600$ spectra from the
SPY survey for weak helium lines, with no new discoveries. However, an
abundance ratio of He:H = 1 or smaller would in general not be
detectable at the typical S/N of that survey.  We also note, that both
helium-rich objects also have metal traces, indicating that some kind
of accretion has happened and the stars are peculiar. Nevertheless,
from the observational point alone, helium pollution could be the
solution.

From the theoretical side this would be very difficult to understand.
The latest asteroseismological study of ZZ~Cetis
\citep{Castanheira.Kepler08} finds a few objects with
a hydrogen layer $\log M_H/M < -8 $ (none $< -10$), but the vast
majority have much thicker layers where no mixing is expected in the
ZZ~Ceti range.

\section{The treatment of convection}
The description of the energy flux by convection is one of the few
remaining problems in stellar astrophysics. In the field of white
dwarfs it has become customary to increase the freedom in the
mixing-length approximation with 3 more parameters in addition to the
mixing-length itself. The different versions are distinguished as ML1,
ML2, ML3, with the mixing length added as in ML1/$\alpha=2.0$ or short
ML1/2.0 \citep[see e.g.][for the nomenclature]
{Fontaine.Villeneuve.ea81, Koester.Allard.ea94}.  For our current
study the main point is that these versions differ in the efficiency
of energy transport: the more efficient a version is the smaller can
the temperature gradient be which transports the necessary flux. This
temperature gradient in turn has a strong influence on the line
profiles, and to a lesser degree also on the continuum energy
distribution.
\begin{figure}[ht]
\begin{minipage}[t]{7.7cm}
\includegraphics[width=7.7cm]{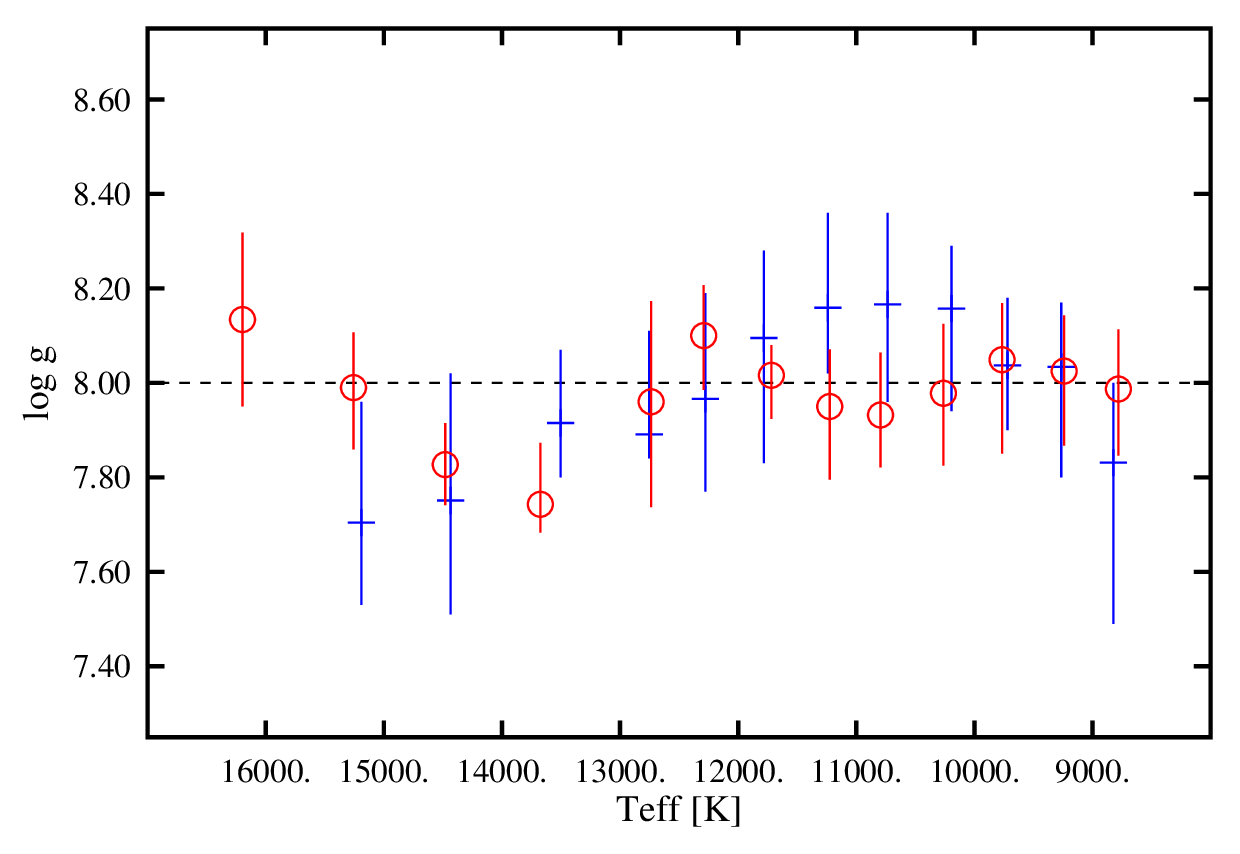}
\caption{\label{conv1} Convection completely suppressed, energy
  transport by radiation only.} 
\end{minipage}\hspace{0.4cm}\begin{minipage}[t]{7.7cm}
\includegraphics[width=7.7cm]{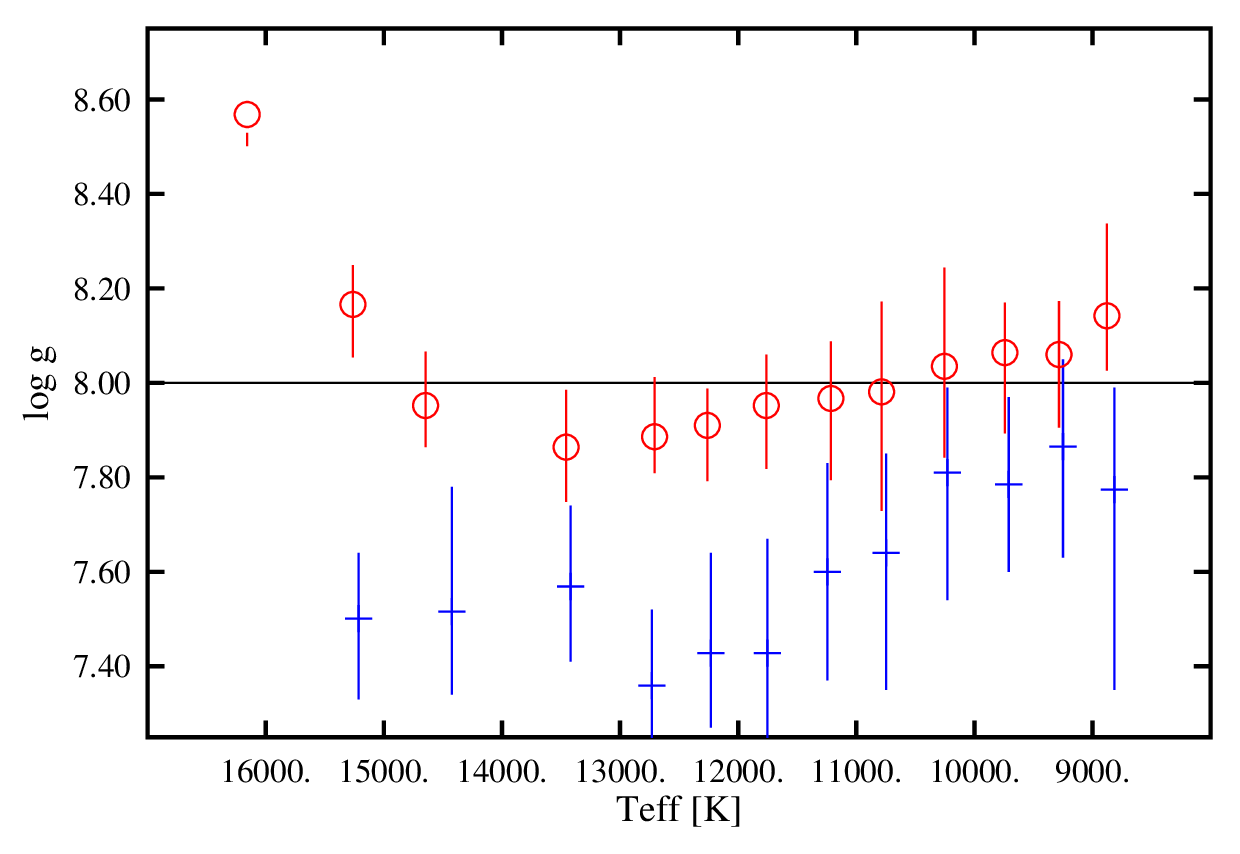}
\caption{\label{conv2} Temperature gradient forced to adiabatic in
  convective regions.}
\end{minipage} 
\end{figure}
We have experimented with various mixing-lengths in the
framework of the ML1 and ML2 versions. The changes always bring changes
to the sequence of average \logg, but none solves the problem in the
sense that a constant average results. As two extreme examples (which
are outside the ML framework) we present in Figs.~\ref{conv1} and
\ref{conv2} a model set with convection totally suppressed (extremely
inefficient) and another one with the convective gradient forced to the
adiabatic gradient (infinite efficiency). In the second case this has
an influence even at rather high temperatures, because there are
thin convection zones, which are, however, normally inefficient and don't
change the radiative gradient.

While this result is not encouraging, we nevertheless want to argue,
that incorrect description of the convective gradient is still the
most likely and in our opinion the only possibility to account for the
observed effect. 

First, there is a very striking coincidence near $\approx 11500$~K
between the steep increase in the thickness of the convection zone and
the increase in the average \logg. Of the other possible solutions
only the helium contamination would be able to explain this
coincidence. 

Second, and even more important: we definitely know that our current
description of convection is incorrect. The current choice of
parameters in the mixing-length approximation (ML2/0.6, or very
similarly ML1/1.75) was introduced by \citet{Koester.Allard.ea94} and
  \citet{Bergeron.Wesemael.ea95} to allow a consistent parameter
determination from optical and ultraviolet spectra. The effective
temperatures found with this version for ZZ~Cetis are in many cases so
hot, that the thermal time scale of the CVZ is at most a few seconds,
much shorter than the pulsation periods. On theoretical grounds, these
timescales are expected to be of similar magnitude, and consequently
\citet{Bergeron.Wesemael.ea95} concluded that no single choice of ML
parameters can simultaneously reproduce optical and UV spectra {\em
  and} the location of the instability strip. Convection must be more
efficient in deeper layers than in the atmosphere, at least when
studied within the ML framework.

This same conclusion was reached completely independently by
\citet{Ludwig.Jordan.ea94}. When comparing the temperature profile
from a two-dimensional hydrodynamic simulation of a white dwarf with
\Teff\ = 12600~K, \logg\ = 8 with mixing-length models of the
atmosphere and envelope they found that the mixing-length must
increase with depth.

\section{Conclusions}
The frustrating result of this study is that we cannot propose any
change of approximations in the input physics, which would clearly
lead to the desired result -- a constant average mass for all white
dwarfs for all effective temperatures. Variations of parameters in the
Hummer-Mihalas-D\"appen occupation probabilities can be
excluded. Helium pollution cannot be definitely excluded, but could be
tested with high resolution, high S/N spectra of (apparent) DAs in the
range 10000 - 12000~K. Because of the asteroseismology results discussed
above we consider this solution unlikely.

That leaves by default convection as the only suspect. Given the
current amount of uncertainty about the correct description, and the
contradictions in the current applications, we consider it quite
possible that a more physically sound description of the convective
flux and gradient might solve the problem. 

At the end of this study a final word of caution: fitting observed
spectra with theoretical models is as much an art as an objective
science. One should be very sceptical of results from spectra with S/N
$<$ 30, as well as DA spectra that do not include the higher Balmer
lines with high S/N. Never believe the errors so readily calculated
by a $\chi^2$ routine; these are errors which consider only
statistical errors in the observed data. With modern CCD spectra from
large telescopes systematic effects of reduction, the fitting process,
and model deficiencies result in much larger errors. These can only be
estimated in a meaningful way from comparing results for the same
objects from different observations, and if possible, from different
authors, with different fitting routines and model grids.

\subsection{Acknowledgments}
D.K. wishes to thank the German DAAD and Brazilian CAPES for a grant
to visit his colleagues in Brazil, as well as for their kind
hospitality in Florianopolis and Porto Alegre.

\end{document}